\documentclass[letterpaper]{article}
\usepackage{aaai2027}
\usepackage[hyphens]{url}
\usepackage{graphicx}
\usepackage{natbib}
\usepackage{caption}
\usepackage{booktabs}
\usepackage{array}
\usepackage{multirow}
\usepackage{amsmath}
\usepackage{amssymb}
\newcommand{\unirag}{UniRAG}
\newcommand{\oursrowpad}{}

\title{DEPT: Document Embedding Preservation Tuning for Unified Query Expansion and Retrieval}

\author{
Jingyuan Wang\textsuperscript{\rm 1},
Richong Zhang\textsuperscript{\rm 1}\corresponding,
Zhijie Nie\textsuperscript{\rm 1},\\
Mingxin Li,
Yanzhao Zhang
}

\affiliations{
\textsuperscript{\rm 1}Beihang University, China\\
wangjy25@buaa.edu.cn, zhangrc@act.buaa.edu.cn
}

\begin{document}

\maketitle

\begin{abstract}
Large language models (LLMs) can both expand underspecified queries and encode text as dense representations, suggesting a unified model for query expansion and retrieval.
Existing systems usually rely on prompted expansions, independently trained modules, or staged optimization, leaving generated expansions only indirectly aligned with the retrieval loss that judges them.
We train a single decoder-only LLM end to end, where the same model generates the expansion and encodes both the expanded query and candidate documents.
This unified setting creates a moving-target problem: retrieval supervision should improve query-side expansion, but the same update also shifts the document embeddings that serve as retrieval targets.
We introduce Document Embedding Preservation Tuning (DEPT), which keeps tuned document embeddings close to cached initial embeddings while allowing retrieval gradients to pass through straight-through decoding into the generator.
DEPT converts joint query--document movement into query-side adaptation against approximately stable, whitened document embeddings that support index reuse and online hard-negative mining.
Experiments with Qwen3-4B-Instruct-2507 and LLaMA-3.2-3B-Instruct on five datasets in BEIR benchmark show that DEPT improves average retrieval quality over training-free, independently trained, and staged unified baselines, while ablations isolate the effects of preservation, whitening, end-to-end expansion training, and online negatives.
Code is available at \url{https://github.com/ILSparkle/DEPT}.
\end{abstract}

\section{Introduction}

Query expansion and dense retrieval solve two complementary parts of retrieval \citep{carpineto2012survey,karpukhin-etal-2020-dense,lin2021pretrained}.
Expansion rewrites an underspecified query into text that exposes missing entities, constraints, or topical cues, while dense retrieval maps the expanded query and documents into a vector space where relevant evidence can be found efficiently.
The combination is attractive because expansion makes the query more explicit and the dense retriever gives the expanded query a scalable search mechanism.
The main obstacle is that these two parts are often optimized through different signals: expansion is judged by textual plausibility or indirect retrieval feedback, whereas the final system is judged by whether the resulting representation ranks the right documents.

\begin{figure}[t]
\centering
\includegraphics[width=0.92\columnwidth]{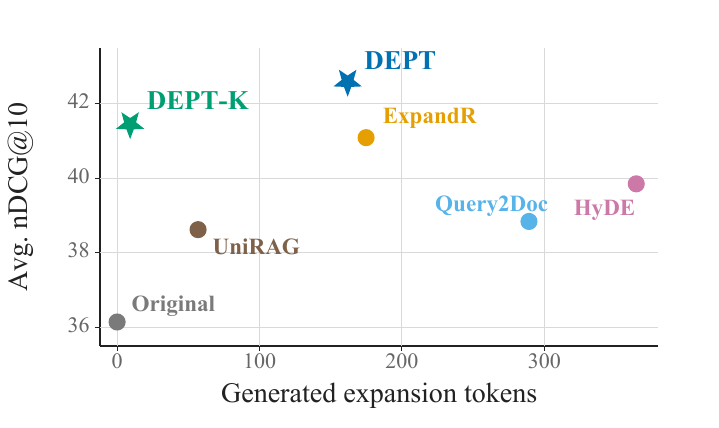}
\caption{Qwen3-4B-Instruct-2507 retrieval quality versus expansion length. Stars mark our methods: \textsc{DEPT} has the best average nDCG@$10$, while \textsc{DEPT-K} gives the short-expansion trade-off.}
\label{fig:quality-cost-qwen}
\end{figure}

\begin{figure*}[t]
\centering
\includegraphics[width=\textwidth]{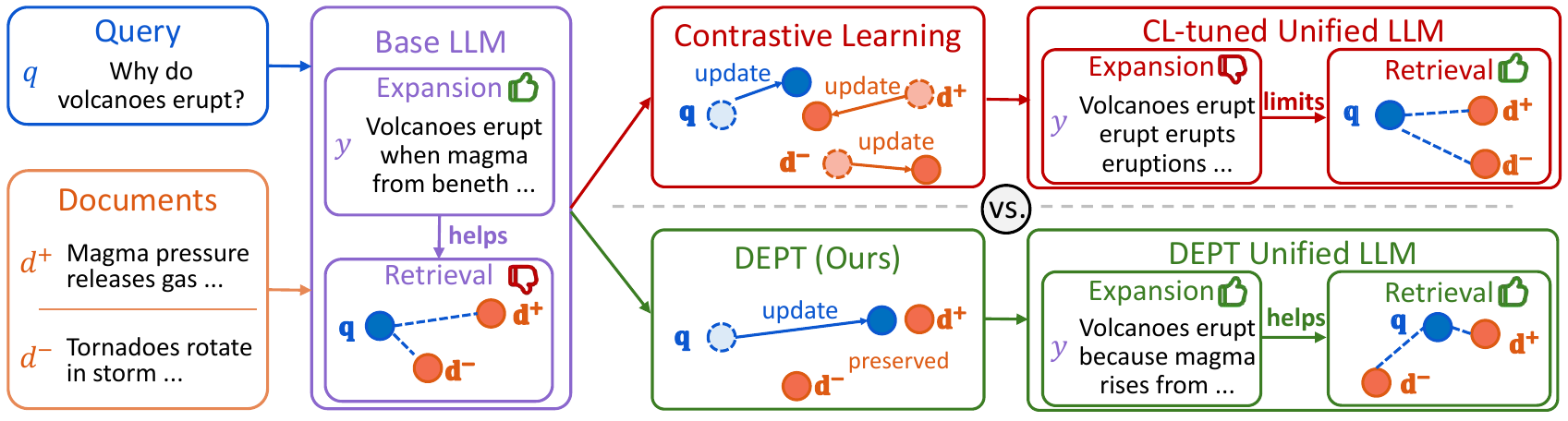}
\caption{Alternative routes for adapting a generative LLM to unified expansion and retrieval. The base model produces useful expansions but weak retrieval representations. Conventional contrastive learning updates both query and document representations, so retrieval improves substantially but degraded expansions can limit the final gain. DEPT instead adapts the query side while preserving document representations, retaining expansion quality and yielding stronger retrieval.}
\label{fig:training-choices}
\end{figure*}

LLMs make this mismatch more visible and more promising to address.
The same model family can produce fluent expansion text and provide dense representations, so one might expect a single decoder-only LLM to learn expansions directly from the retrieval loss that evaluates them.
Such a unified model would have two practical advantages.
First, expansion would become a trainable retrieval action rather than a prompted preprocessing step.
Second, the generated text would remain inspectable, so the retrieval decision could still be examined through the model's textual output rather than only through an opaque vector.

Existing LLM-based expansion systems move toward this goal but stop short of the end-to-end objective.
Training-free methods prompt an LLM to produce pseudo-documents or query augmentations, leaving the generator unchanged even when the downstream retriever fails \citep{gao-etal-2023-precise,wang-etal-2023-query2doc}.
Independent-training methods such as InPars \citep{bonifacio2022inpars} and Promptagator \citep{dai2023promptagator} improve the training signal, but the generator and retriever are still optimized through separate interfaces.
Sequential systems use retrieval feedback or preference-style supervision, as in ExpandR \citep{yao-etal-2025-expandr}, yet the generator is not optimized by the final contrastive loss of the deployed retriever.
Staged unified methods such as \unirag{} \citep{li-etal-2025-unirag} reduce architectural separation, but augmentation and representation remain separate training phases.
The common limitation is therefore not that generated text is useless; it is that the text-producing component is not continuously shaped by the same retrieval objective and embedding geometry that determine final ranking.

Directly training the unified model, however, creates a different failure mode.
Figure~\ref{fig:training-choices} illustrates the choice.
Before retrieval training, a base LLM can already produce useful expansions, but its dense retrieval representations are weak.
Conventional contrastive learning fixes the representation problem by moving expanded queries toward relevant documents and away from irrelevant ones.
Because the same parameters also define document embeddings and generation behavior, the update can improve ranking while changing the retrieval targets and the expansion policy at the same time.
The outcome is a bottleneck rather than a complete failure: retrieval still improves, but the final gain is limited when expansion drifts away from the fluent explanatory behavior that made LLM augmentation useful in the first place.

This failure mode suggests that unified expansion and retrieval should be asymmetric.
The query side should remain plastic because the generated tokens and expanded-query representation are the parts that need to learn from retrieval feedback.
The document side should instead serve as a stable target because it defines the contrastive comparison during training.
As a practical benefit, the same stability can also support cached indexing at inference.
With relatively fixed document targets, the retrieval objective places clearer pressure on the query path and expansion behavior.
Without that stability, the same objective can be satisfied by moving documents rather than by improving the generated query.

We introduce \textbf{Document Embedding Preservation Tuning (DEPT)} to implement this asymmetric training principle.
DEPT keeps current document embeddings close to cached embeddings from the initial model, while allowing retrieval gradients to update expansion and the expanded-query representation.
This preservation makes two additional design choices useful rather than fragile.
Since raw LLM embeddings can be anisotropic and poorly calibrated for cosine retrieval \citep{ethayarajh-2019-contextual}, we fit a fixed whitening transform on cached document embeddings and train in the resulting coordinate system.
Because document embeddings remain close to this reference, the same cached index can support online hard-negative mining during training and index reuse at inference.
Figure~\ref{fig:quality-cost-qwen} previews the resulting trade-off: the long-expansion model reaches the best average retrieval quality, while the keyword-style variant remains competitive with very short expansions.

The main contributions of this work can be summarized as follows.
First, we formulate query expansion and dense retrieval as a unified decoder-only LLM training problem, and identify document-embedding drift as the stability obstacle that prevents ordinary contrastive learning from fully exploiting expansion.
Second, we propose Document Embedding Preservation Tuning (DEPT), which uses a Document Embedding Preservation (DEP) loss, fixed whitening, online hard-negative mining, and straight-through expansion training to make retrieval gradients act primarily on the query side.
Third, experiments on two LLM backbones and five retrieval tasks show that DEPT achieves state-of-the-art average retrieval quality among the compared query-expansion paradigms while preserving generation ability and cached-index compatibility.

\begin{figure*}[t]
\centering
\ifdefined\PLACEHOLDEROVERVIEW
\fbox{\begin{minipage}[c][0.36\textheight][c]{0.96\textwidth}
\centering
\Large Figure 2 Placeholder\\[0.75em]
\normalsize Main overview figure to be redesigned.\\[0.35em]
\small Planned flow: unified expansion and retrieval $\rightarrow$ end-to-end retrieval gradients $\rightarrow$ straight-through expansion, document embedding preservation, and fixed whitening.
\end{minipage}}
\else
\includegraphics[width=\textwidth]{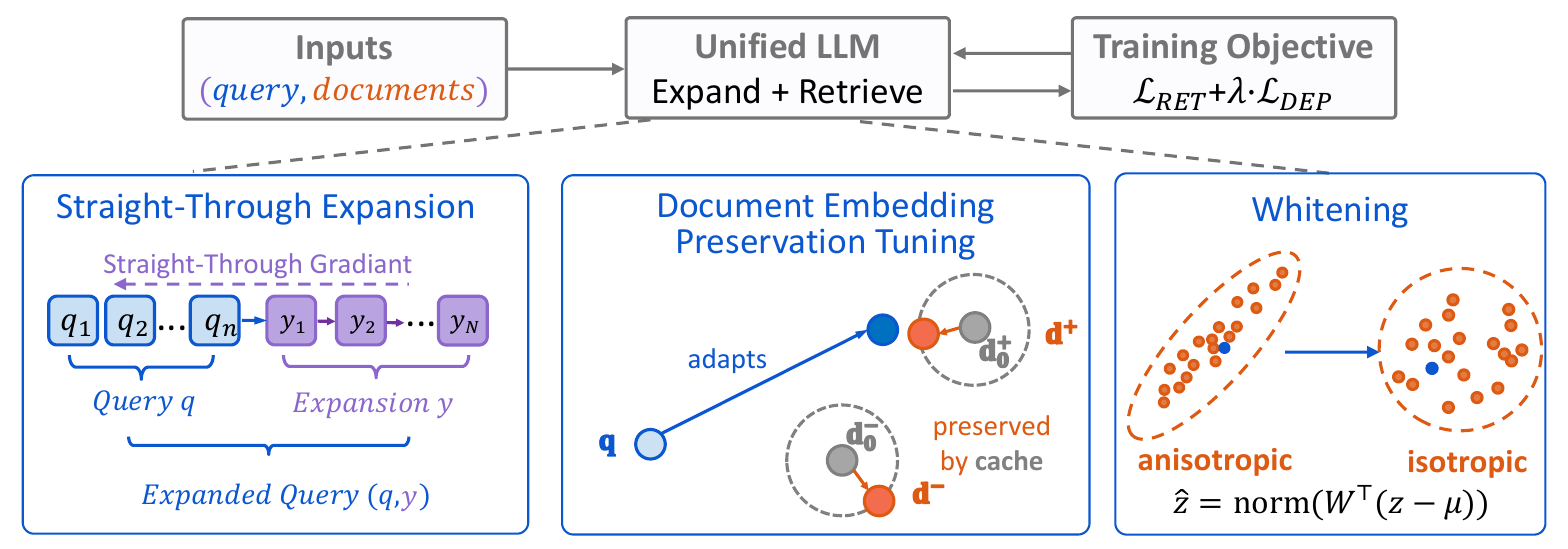}
\fi
\caption{Overview of DEPT training. A single decoder-only LLM handles expansion and retrieval under an end-to-end retrieval objective. The lower panels detail the three mechanisms: straight-through expansion passes retrieval gradients to generated tokens, the DEP loss preserves current document embeddings near cached document embeddings from the initial model, and a fixed whitening transform maps embeddings into a better-conditioned retrieval space.}
\label{fig:overview}
\end{figure*}

\section{Related Work}

\paragraph{LLM-Based Query Expansion.}
HyDE encodes a generated hypothetical document to bridge a zero-shot query--document gap \citep{gao-etal-2023-precise}, while Query2Doc prompts an LLM to generate a pseudo-document that augments the original query \citep{wang-etal-2023-query2doc}.
ExpandR uses LLM guidance to train a retriever beyond the literal query and aligns generated augmentations with retrieval preferences \citep{yao-etal-2025-expandr}.
\unirag{} is the closest staged unified framework: it uses a decoder-only LLM for both query augmentation and representation, but optimizes augmentation and encoding through separate stages \citep{li-etal-2025-unirag}.
These methods show that generated text can improve retrieval, but generation is usually optimized outside the final contrastive retrieval loss or through indirect feedback.
They leave open how to stabilize retrieval targets when one shared model handles generation and encoding.

\paragraph{Generative and Representational LLMs.}
Instruction-tuned LLM embeddings show that generative models can produce strong text representations \citep{wang-etal-2024-improving-text}.
GritLM jointly trains generative and embedding objectives in a single model \citep{muennighoff2024gritlm}, establishing the broader feasibility of unifying generation and representation.
Our setting adds a retrieval-specific constraint that is less visible in generic multitask formulations: corpus embeddings are retrieval targets during training and can be expensive to refresh in deployment.
With one model for generation and embedding, shared updates can change both query behavior and document embeddings, so end-to-end expansion training must control target drift rather than treating document movement as harmless.

\paragraph{Representation Geometry and Stability.}
Transformer representations are known to be anisotropic \citep{ethayarajh-2019-contextual}, and whitening can improve semantic similarity and retrieval by centering and decorrelating sentence embeddings \citep{su2021whitening}.
Whitening is commonly used as a post-processing transform for a fixed embedding model.
In trainable retrieval systems, the transformed space remains valid only if embeddings stay close to the distribution used to estimate the transform.

\section{Method}

\subsection{Preliminary}

We first describe the ordinary expansion-and-retrieval pipeline before introducing training.
Given a query $q_i$, an expansion module generates a text sequence $y_i=(y_{i,1},\ldots,y_{i,T})$, where $T$ is the expansion length.
The retriever then encodes the concatenated text $[q_i;y_i]$ and each document $d$ from a corpus $\mathcal{D}$ into a shared vector space, and ranks documents by similarity to the expanded query.

Let $\theta$ denote the parameters of the encoder used by the retriever.
For any token sequence $x$, the model produces final-layer hidden states $H(x)=(h_1(x),\ldots,h_{|x|}(x))$.
We use one raw embedding function for both sides, defined by mean pooling followed by $\ell_2$ normalization:
\begin{equation}
\begin{aligned}
\bar h(x)
&=\frac{1}{|x|}\sum_{t=1}^{|x|}h_t(x), \\
f_\theta(x)
&=\frac{\bar h(x)}{\|\bar h(x)\|_2}.
\end{aligned}
\end{equation}
Given a generated expansion $y_i$, we abbreviate the raw expanded-query embedding and the raw embedding of document $d$ as
\begin{equation}
\mathbf{q}_i=f_\theta([q_i;y_i]),
\qquad
\mathbf{d}=f_\theta(d),
\end{equation}
where $[q_i;y_i]$ denotes concatenation of the original query and its expansion.
At inference, retrieval uses a score such as $\mathbf{q}_i^\top\mathbf{d}$ to rank documents.
Existing methods differ mainly in how the expansion module and retriever are obtained: the expansion may be prompted, trained separately, or trained in a stage before the final retriever.
In all cases, the basic interface is the same: generated text changes the query representation, and the retriever scores it against document embeddings.
Figure~\ref{fig:overview} summarizes the resulting training graph and highlights the three mechanisms used by DEPT: straight-through expansion, document embedding preservation, and fixed whitening.

\subsection{Document Embedding Preservation Tuning}

DEPT uses one decoder-only LLM for both abilities in the pipeline: it generates the expansion and also encodes expanded queries and documents.
This single-model formulation is important because the expansion and representation are expressed by the same parameters, but it also creates a conflict between two roles of the model.
The query side should change so that generated expansions become better retrieval actions.
The document side should remain stable because it defines the retrieval targets used by the contrastive loss; cached-index reuse is a downstream benefit.

This conflict is asymmetric rather than a generic preference for small updates.
The query-side generator is conditioned on a retrieval instruction and the input query, and the query representation is computed from the expanded input $[q_i;y_i]$.
This path should be allowed to adapt, because retrieval supervision must teach the model what kind of expansion improves document ranking.
In contrast, documents are fed to the same model as plain document text, and their embeddings serve as the retrieval targets against which all expanded queries are judged.

This asymmetry also helps preserve generation: plain-document encoding shares the same token embeddings and transformer blocks used by document-conditioned language modeling, so keeping document embeddings close to their cached initial embeddings acts as lightweight functional rehearsal while the instruction-conditioned query path remains free to adapt.

The conflict appears directly in the raw query--document score
$r(q_i,y_i,d)=\mathbf{q}_i^\top\mathbf{d}$.
Because query and document embeddings share parameters, this score has the gradient
\begin{equation}
\begin{aligned}
\nabla_\theta r(q_i,y_i,d)
&=
\left(\nabla_\theta \mathbf{q}_i\right)^\top \mathbf{d} \\
&\quad
+ \mathbf{q}_i^\top
\left(\nabla_\theta \mathbf{d}\right).
\end{aligned}
\end{equation}
The first term is the desired query-side update: it teaches the expansion and expanded-query representation to align with relevant documents.
The second term moves the document representation, creating query--document gradient interference.
If both sides are left unconstrained, the model can reduce retrieval loss by reorganizing document embeddings around the current minibatch rather than learning expansions that transfer to a stable retrieval target.
DEPT resolves this single-model, dual-ability conflict by keeping document embeddings close to their cached references while leaving the query path trainable.
This makes query optimization more consistent across steps: target document embeddings change slowly, so retrieval gradients are directed toward improving the expansion and query representation instead of chasing a moving corpus representation.

Freezing a separate document encoder would stabilize targets but break the unified formulation.
We instead regularize the quantity that matters to retrieval: the raw document embedding $\mathbf{d}$ produced from each document $d$.
Let $\theta_0$ denote the initial model before DEPT tuning.
Before training, we precompute a cached reference embedding for each training document:
\begin{equation}
\mathbf{d}^0=f_{\theta_0}(d).
\end{equation}
These cached embeddings are not updated during training.

For any training document $d$, define its angular drift from the cached reference embedding as
\begin{equation}
\delta(d)=1-\cos\!\left(\mathbf{d},\mathbf{d}^0\right).
\end{equation}
Controlling this drift keeps retrieval feedback tied to an adapting expanded query rather than to arbitrarily moving target document embeddings.
In a training step with $B$ queries, let $\mathcal{C}_i$ be the candidate documents for query $q_i$, and let $\mathcal{S}_i\subseteq\mathcal{C}_i$ be the documents whose current embeddings are computed, including the positive document and sampled negatives.
For scale $s>0$ and exponent $p>1$, we define the DEP loss as
\begin{equation}
\mathcal{L}_{\mathrm{DEP}}=
\frac{1}{B}\sum_{i=1}^{B}\frac{1}{|\mathcal{S}_i|}
\sum_{d\in\mathcal{S}_i}
\left(s\,\delta(d)\right)^p .
\end{equation}
Here $\lambda>0$ will control the preservation strength in the final objective.
The scale $s$ keeps small cosine deviations numerically visible, while $p>1$ emphasizes large departures.
Reference document embeddings $\mathbf{d}^0$ receive no gradient.

\begin{figure}[t]
\centering
\includegraphics[width=\columnwidth]{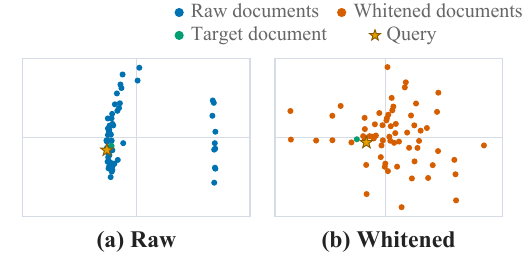}
\caption{Schematic illustration of whitening, visualized by projecting embeddings onto the top two principal components via PCA. Raw cached document embeddings can concentrate along dominant directions; the fixed whitening transform maps them into a better-conditioned retrieval space.}
\label{fig:whitening}
\end{figure}

\paragraph{Whitening Preserved Document Embeddings.}
The DEP loss preserves raw document embeddings, but retrieval still benefits from a better-conditioned coordinate system.
Because decoder-only LLM embeddings can be anisotropic, using the raw embedding geometry inherits its defects.
Figure~\ref{fig:whitening} illustrates how whitening reduces dominant directional variance before cosine retrieval.

We therefore estimate a fixed whitening transform from cached reference document embeddings and keep this transform fixed throughout tuning.
This design is tied to preservation: whitening improves the structure of the initial document embedding distribution, and the DEP loss keeps later document embeddings close enough for the same coordinate system to remain useful.
For independently trained baselines whose document embeddings continue to move, a whitening transform fitted before training would gradually cease to describe the current embedding distribution and would behave mainly as an extra fixed linear map.
Let $\mu$ be the empirical mean of cached document embeddings $\mathbf{d}^0$, and let $U\Lambda U^\top$ be the eigendecomposition of their covariance matrix.
With small constant $\varepsilon>0$ and whitening strength $\alpha\in[0,1]$, we define
\begin{equation}
W = U(\Lambda+\varepsilon I)^{-\alpha}.
\end{equation}
Here $I$ is the identity matrix, and $\varepsilon$ prevents unstable inversion of small eigenvalues.
Any raw embedding $z$ is mapped into the fixed whitened coordinate system by
\begin{equation}
\hat z=\operatorname{norm}\!\left(W^\top(z-\mu)\right).
\end{equation}
We write $\hat{\mathbf{q}}_i$ for the whitened version of $\mathbf{q}_i$, and $\hat{\mathbf{d}}$ for the whitened version of $\mathbf{d}$.
Since $(\mu,W)$ is fitted from cached documents before tuning, whitening remains meaningful only if the DEP loss keeps current document embeddings close to their cached references.

\subsection{End-to-End Expansion and Retrieval Training}

With stable document embeddings in place, the retrieval objective can be used to train expansion and representation jointly.
The role of DEPT is to make this end-to-end signal selective: retrieval gradients may update expansion logits and the expanded-query representation, while the DEP loss prevents the same updates from turning retrieval into a moving-target problem on the document side.
The training set contains triples $(q_i,d_i^+,\mathcal{D}_i^-)$, where $q_i$ is the input query, $d_i^+$ is a relevant document, and $\mathcal{D}_i^-$ is a pool of non-relevant or mined negative documents.
For a minibatch of size $B$, the candidate set $\mathcal{C}_i$ contains $d_i^+$, in-batch documents, and sampled negatives.
The shared model first generates $y_i$ from the retrieval instruction and query, then encodes $[q_i;y_i]$ and every document in $\mathcal{C}_i$.

Hard token selection would sever the path between the retrieval loss and the expansion logits.
We retain discrete expansions in the forward pass while using a differentiable approximation in the backward pass.
For one generation step, let $\ell(v)$ be the logit of vocabulary token $v\in\mathcal{V}$, let $\mathcal{K}$ be the top-$k$ tokens under that logit vector, and let $E(v)$ be the embedding of token $v$.
We compute
\begin{align}
\pi(v) & =
\frac{\exp(\ell(v))}
{\sum_{u\in\mathcal{K}}\exp(\ell(u))},
\qquad v\in\mathcal{K}, \\
\widetilde e & =
\sum_{v\in\mathcal{K}}\pi(v)E(v), \\
e^{\mathrm{ST}} & =
\widetilde e
+ \operatorname{sg}\!\left(E(\widehat y)-\widetilde e\right),
\end{align}
where $\widehat y=\arg\max_{v\in\mathcal{V}}\ell(v)$ is the greedy token and $\operatorname{sg}(\cdot)$ stops gradients.
Thus, $e^{\mathrm{ST}}$ equals the greedy token embedding in the forward computation, while its gradient is that of the top-$k$ soft embedding.
Applying this step at each generation position gives $(e_{i,1}^{\mathrm{ST}},\ldots,e_{i,T}^{\mathrm{ST}})$.
The query encoder consumes these expansion embeddings together with the original query, allowing retrieval gradients to update both the representation and the logits that produced the expansion.
This path trains the unified generator--encoder interface: an expansion is useful only if its embedding ranks the relevant document above candidates in $\mathcal{C}_i$.

The retrieval loss is the InfoNCE objective in this whitened space:
\begin{equation}
\mathcal{L}_{\mathrm{RET}} = -\frac{1}{B}\sum_{i=1}^{B}
\log
\frac{
\exp\!\left(\hat{\mathbf{q}}_i^\top
\hat{\mathbf{d}}_i^+/\tau\right)}
{\sum_{d\in\mathcal{C}_i}
\exp\!\left(\hat{\mathbf{q}}_i^\top
\hat{\mathbf{d}}/\tau\right)} ,
\end{equation}
where $\hat{\mathbf{d}}_i^+$ is the whitened embedding of the positive document $d_i^+$.
The final DEPT objective is
\begin{equation}
\mathcal{L}_{\mathrm{DEPT}}=
\mathcal{L}_{\mathrm{RET}}+\lambda\mathcal{L}_{\mathrm{DEP}}.
\end{equation}
Here $\mathcal{L}_{\mathrm{RET}}$ supplies the end-to-end learning signal, while $\mathcal{L}_{\mathrm{DEP}}$ keeps that signal focused on query-side expansion and representation learning rather than arbitrary corpus movement.
The loss does not match query embeddings to their initialization or freeze parameters; retrieval gradients still traverse query encoding, expansion, and any parameters whose updates keep document embeddings close to their cached references.

\subsection{Leveraging Stable Document Embeddings}

\paragraph{Index Reuse.}
Dense retrieval normally encodes the corpus once and stores its vectors in a nearest-neighbor index \citep{karpukhin-etal-2020-dense,xiong2021approximate}.
End-to-end tuning without preservation breaks this serving assumption: even if the tuned model improves retrieval after re-encoding all documents, the index built before tuning may no longer contain vectors compatible with the tuned queries.
DEPT is designed to avoid this mismatch.
Since document embeddings are kept close to their cached references and whitening is fitted on the same cached document embeddings, the tuned expanded-query embedding can be searched against the reference document index without rebuilding the corpus.
This gives a direct deployment benefit for large collections, where document encoding and index construction are often much more expensive than updating the query-side model.
We evaluate this compatibility by comparing retrieval with the cached reference index against retrieval after re-encoding documents with the final model.

\paragraph{Online Hard-Negative Mining.}
The stable index also supports training-time hard-negative mining.
At each step, the current model produces an expansion and a whitened query embedding $\hat{\mathbf{q}}_i$.
We use this query to search a fixed FAISS index built from cached whitened document embeddings, remove the positive document, and sample negatives from a high-ranking interval.
The selected negative documents are then re-encoded by the current model before entering the InfoNCE denominator and the DEP loss.
This procedure gives the retrieval loss query-conditioned negatives that track the current expansion behavior, while the expensive corpus index remains fixed.
The sampling interval also avoids relying only on the nearest retrieved items, which can include mislabeled positives or overly ambiguous documents, and provides a controlled source of difficult but usable negatives throughout training.

\begin{table*}[!t]
\centering
\footnotesize
\setlength{\tabcolsep}{1mm}
\begin{tabular*}{\textwidth}{@{\extracolsep{\fill}}llrcccccc@{}}
\toprule
Backbone & Method & \multicolumn{1}{r}{Exp. Tok.} & SciFact & ArguAna & NFCorpus & FiQA & SCIDOCS & Avg. \\
\midrule
\multirow{7}{*}{\shortstack[l]{Qwen3-4B\\Instruct-2507}}
& Original Query & 0.00 & 66.26 & 32.68 & 31.34 & 32.92 & 17.48 & 36.14 \\
& Query2Doc & 289.18 & 72.28 & 31.04 & 34.28 & 37.09 & \underline{19.51} & 38.84 \\
& HyDE & 364.50 & 72.16 & 38.91 & 34.16 & 35.16 & 18.86 & 39.85 \\
& ExpandR & 174.82 & 72.66 & \underline{40.74} & 35.62 & \underline{37.41} & 19.02 & 41.09 \\
& \unirag{} & 56.86 & 67.28 & 39.08 & 33.54 & 35.35 & 17.83 & 38.62 \\
& \oursrowpad\textsc{DEPT-K} \textbf{(Ours)} & 9.15 & \underline{73.62} & 40.30 & \underline{35.81} & 37.36 & \textbf{20.14} & \underline{41.45} \\
& \textsc{DEPT} \textbf{(Ours)} & 161.87 & \textbf{74.35} & \textbf{41.12} & \textbf{36.99} & \textbf{41.66} & 18.82 & \textbf{42.59} \\
\midrule
\multirow{7}{*}{\shortstack[l]{LLaMA-3.2\\3B-Instruct}}
& Original Query & 0.00 & 65.85 & 33.98 & 32.55 & 31.94 & 16.60 & 36.18 \\
& Query2Doc & 262.51 & 69.48 & 34.01 & 31.22 & \underline{34.75} & \textbf{19.21} & 37.73 \\
& HyDE & 307.38 & 65.30 & 34.84 & 26.89 & 26.15 & 15.61 & 33.76 \\
& ExpandR & 164.86 & 69.28 & \underline{38.02} & 32.40 & 34.17 & 16.96 & 38.17 \\
& \unirag{} & 59.18 & 67.73 & 35.61 & 30.29 & 33.57 & 16.35 & 36.71 \\
& \oursrowpad\textsc{DEPT-K} \textbf{(Ours)} & 8.98 & \underline{69.83} & 37.14 & \underline{34.45} & 34.62 & \underline{17.63} & \underline{38.73} \\
& \textsc{DEPT} \textbf{(Ours)} & 148.82 & \textbf{71.36} & \textbf{38.56} & \textbf{35.22} & \textbf{35.53} & 17.31 & \textbf{39.60} \\
\bottomrule
\end{tabular*}
\caption{Main retrieval results on five BEIR tasks, reported as nDCG@$10$. Exp. Tok. is the average number of generated expansion tokens. \unirag{} denotes our reimplementation under the same backbone family and retrieval protocol. Within each backbone, the best result is bold and the second-best is underlined.}
\label{tab:main-results}
\end{table*}

\section{Experiments}

\subsection{Experimental Setup}

\paragraph{Training Data.}
We train on an ECHO retrieval mixture constructed from eight sources: ELI5 question answering \citep{fan-etal-2019-eli5}, FEVER \citep{thorne-etal-2018-fever}, HotpotQA \citep{yang-etal-2018-hotpotqa}, MS~MARCO document retrieval, MS~MARCO passage retrieval \citep{nguyen2016msmarco}, Natural Questions \citep{kwiatkowski-etal-2019-natural}, SQuAD \citep{rajpurkar-etal-2016-squad}, and TriviaQA \citep{joshi-etal-2017-triviaqa}.
Each example is converted to an instruction, query, positive document, and negative document in an E5-style format \citep{wang2022text}.
The expansion prompt contains the retrieval instruction and original query; we use a keyword-style prompt for \textsc{DEPT-K} and a longer explanatory prompt for \textsc{DEPT}.

\paragraph{Evaluation Tasks and Metric.}
We evaluate zero-shot retrieval on five datasets in BEIR benchmark \citep{thakur2021beir}: SciFact \citep{wadden-etal-2020-fact}, ArguAna \citep{wachsmuth-etal-2018-retrieval}, NFCorpus \citep{boteva2016nfcorpus}, FiQA \citep{maia2018www}, and SCIDOCS \citep{cohan-etal-2020-specter}.
We report nDCG@$10$ \citep{jarvelin2002cumulated} for every task and the unweighted average across tasks.

\paragraph{Backbones and Baselines.}
We instantiate every method with Qwen3-4B-Instruct-2507 and LLaMA-3.2-3B-Instruct.
Training-free baselines include the original query, HyDE \citep{gao-etal-2023-precise}, and Query2Doc \citep{wang-etal-2023-query2doc}; HyDE and Query2Doc use the untrained instruction model only for expansion and the corresponding contrastive-tuned LLM retriever for encoding.
ExpandR \citep{yao-etal-2025-expandr} represents independent expansion-and-retrieval training with separate same-backbone decoder and encoder instances.
We also include \unirag{} as a staged unified baseline, using our same-backbone reimplementation in which a distilled augmenter supplies text to a subsequently trained retriever but receives no final retrieval gradient.
All methods use the same evaluation collections and metric.

\paragraph{Implementation Details.}
We use mean pooling and $\ell_2$ normalization for all raw embeddings.
\textsc{DEPT-K} uses a keywords-style prompt with a max-32 output limit, while \textsc{DEPT} uses a passage-style prompt with a 128--512-token output range; realized expansion lengths are in Table~\ref{tab:main-results}.
Table~\ref{tab:hyperparams} summarizes the main hyperparameters.
Unless stated otherwise, all reported results are from a single run, and preprocessing, sampling, and training use seed 42.
Training uses LoRA adaptation \citep{hu2022lora} and is conducted on four NVIDIA RTX PRO 6000 GPUs with 96GB memory per GPU.

\begin{table}[t]
\centering
\small
\setlength{\tabcolsep}{1mm}
\begin{tabular}{lcc}
\toprule
Hyperparameter & \textsc{DEPT-K} & \textsc{DEPT} \\
\midrule
Expansion prompt & Keywords-style & Passage-style \\
Output limit & Max 32 tokens & 128--512 tokens \\
Input length & \multicolumn{2}{c}{1,024 tokens} \\
ST decoding & \multicolumn{2}{c}{Top-$64$} \\
LoRA & \multicolumn{2}{c}{$r=16$, $\alpha=32$, dropout $0.05$} \\
Training & \multicolumn{2}{c}{1k steps, lr $10^{-4}$, batch 256} \\
Retrieval temp. & \multicolumn{2}{c}{$0.02$} \\
DEP loss & \multicolumn{2}{c}{$\lambda=0.1$, $s=100$, $p=2$} \\
Whitening & \multicolumn{2}{c}{1,024 cached docs, $\alpha=0.5$} \\
Online mining & \multicolumn{2}{c}{Rank-5 negative} \\
\bottomrule
\end{tabular}
\caption{Main hyperparameters for DEPT variants. Shared values use merged cells.}
\label{tab:hyperparams}
\end{table}

\subsection{Results and Analysis}

The experiments examine five aspects: main retrieval quality, component causality, cached-index compatibility, general generation behavior, and qualitative expansion behavior.
We distinguish \textsc{DEPT-K}, which generates short keyword-style expansions, from long-expansion \textsc{DEPT}.

\paragraph{Main Results.}
Table~\ref{tab:main-results} evaluates whether expansion improves when retrieval supervision trains the generator and encoder in one unified objective rather than through prompting, independent objectives, or staged unification.
Long-expansion \textsc{DEPT} obtains the best average score with both backbones, while \textsc{DEPT-K} remains competitive using roughly nine generated tokens.
Figure~\ref{fig:quality-cost-qwen} shows the same trade-off on Qwen: \textsc{DEPT} is the highest-quality point, and \textsc{DEPT-K} lies on the short-expansion frontier.

The gains are not a simple length effect.
Training-free methods generate longer text, but it is not optimized for the embedding-based scorer that ranks it.
\textsc{DEPT} lets the final retrieval loss update the expansion path inside the shared model; SCIDOCS remains the boundary case, where concise topical cues can be stronger than long semantic expansions.

\paragraph{Ablation Study.}
Table~\ref{tab:ablation} isolates DEPT components on Qwen.
Removing the DEP loss hurts even when documents are re-encoded, showing that preservation improves optimization beyond cached-index reuse.
Whitening is the largest individual factor, and removing expansion or detaching it as an offline input loses the benefit of straight-through training.
Overall, the DEP loss and whitening stabilize document embeddings, while expansion training and hard negatives teach the query side to use those stable targets.

\begin{center}
\centering
\footnotesize
\setlength{\tabcolsep}{1mm}
\begin{tabular}{@{}lcccc@{}}
\toprule
Variant & SciFact & ArguAna & NFCorpus & Avg. \\
\midrule
\textsc{DEPT} \textbf{(Ours)} & \textbf{74.35} & \textbf{41.12} & \textbf{36.99} & \textbf{50.82} \\
w/o DEP Loss & 71.88 & 35.75 & 33.21 & 46.95 \\
w/o Whitening & 66.31 & 30.58 & 29.37 & 42.09 \\
w/o Online Neg. & 72.79 & 37.81 & 34.20 & 48.27 \\
w/ Detached Expansion & 72.33 & 37.25 & 32.37 & 47.32 \\
w/o Expansion & 65.35 & 31.94 & 31.03 & 42.77 \\
\bottomrule
\end{tabular}
\captionof{table}{Ablations on three BEIR tasks with Qwen3-4B-Instruct-2507, reported as nDCG@$10$.}
\label{tab:ablation}
\end{center}

\begin{figure}[t]
\centering
\includegraphics[width=\columnwidth]{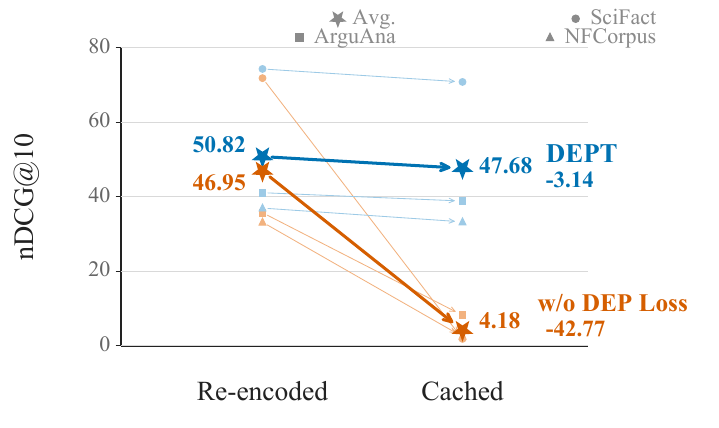}
\caption{Cached-index compatibility on Qwen3-4B-Instruct-2507. Arrows show movement from re-encoded to cached-index retrieval; thick arrows are averages and pale arrows are individual datasets.}
\label{fig:index-reuse-arrows}
\end{figure}
\FloatBarrier

\paragraph{Index Reuse.}
Figure~\ref{fig:index-reuse-arrows} tests whether a pre-tuning document index remains usable after end-to-end training.
With \textsc{DEPT}, cached-index retrieval stays close to re-encoding all documents with the final model.
Without the DEP loss, the original index no longer contains valid keys for the tuned document encoder; preservation prevents this failure mode while allowing the query path and LoRA parameters to adapt.

\begin{figure}[t]
\centering
\includegraphics[width=\columnwidth]{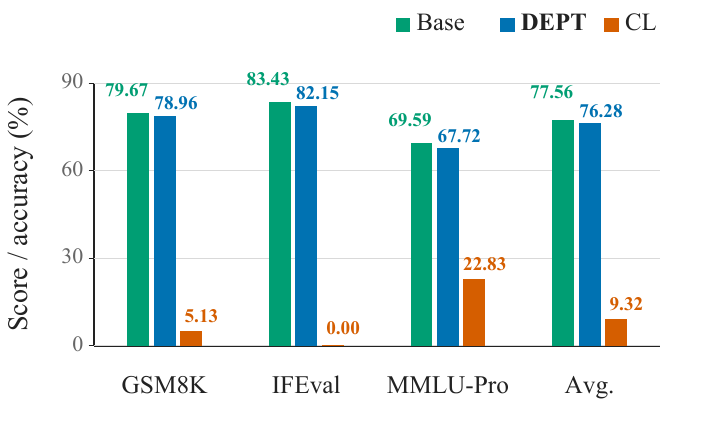}
\caption{General generation evaluation on GSM8K, IFEval, MMLU-Pro, and their average. CL denotes standard contrastive learning on the unified model.}
\label{fig:generation-ability}
\end{figure}

\paragraph{Generation Ability.}
Figure~\ref{fig:generation-ability} checks generation on three benchmarks \citep{cobbe2021training,zhou2023instruction,wang2024mmlupro}.
\textsc{DEPT} stays close to the base model, decreasing from 77.56 to 76.28 on average, whereas standard contrastive learning (CL) collapses to 9.32.
Thus preservation improves retrieval without turning the same decoder into a narrow embedding-only model.

\paragraph{Case Study.}
Table~\ref{tab:case-study} illustrates the expansion behavior behind the aggregate trends.
The base model mentions relevant biomedical mechanisms but does not select them as retrieval cues, whereas standard CL and \unirag{} produce degraded or repetitive strings.
\textsc{DEPT-K} compresses the query into high-value cues, and full \textsc{DEPT} preserves the same mechanism in a readable explanatory form.
This pattern matches the quantitative results: preservation lets the model improve retrieval without discarding the expansion behavior that makes LLM-based retrieval inspectable.

\begin{center}
\centering
\footnotesize
\setlength{\tabcolsep}{1mm}
\begin{tabular}{@{}>{\raggedright\arraybackslash}p{0.16\columnwidth}>{\raggedright\arraybackslash}p{0.76\columnwidth}@{}}
\toprule
\multicolumn{2}{@{}p{0.96\columnwidth}@{}}{\textbf{Query:} Starving Tumors of Their Blood Supply} \\
\midrule
Method & Generated expansion excerpt \\
\midrule
Base & ... cutting off the \textbf{nutrients and oxygen} that tumors need ... \textbf{anti-angiogenic drugs} ... \textbf{vascular endothelial growth factor} ... \\
\textsc{CL} & \textbf{keywords}, Starving Task, Task, \textbf{Starvation}, StarvStarving \\
\unirag{} & ... \textbf{Starving tumors} Starving Starving tumor Star Star Star tumors ... \\
\textsc{DEPT-K} & \textbf{Starving tumors blood supply}; \textbf{tumor hypoxia mechanism}; \textbf{angiogenesis inhibition} \\
\textsc{DEPT} & ... deprive tumors of the \textbf{oxygen and nutrients} they need ... \textbf{anti-angiogenic drugs} ... \textbf{angiogenesis} ... \\
\bottomrule
\end{tabular}
\captionof{table}{Expansion excerpts on one NFCorpus query; three periods mark omitted text.}
\label{tab:case-study}
\end{center}

\section{Conclusion}

We studied unified LLM-based query expansion and dense retrieval, where the same decoder-only model generates retrieval-oriented expansions and encodes the candidate documents that judge them.
The central issue is a moving-target training dynamic: ordinary contrastive updates can improve local retrieval scores while drifting document embeddings and degrading the generator that produces the expansion.
DEPT addresses this problem with document embedding preservation, fixed whitening, online hard negatives, and straight-through expansion training.
Across two LLM backbones and five BEIR tasks, it improves retrieval while preserving generation and cached-index compatibility, showing that unified generative and representational retrieval benefits from explicitly stabilizing document embeddings.
This points to a simple design principle: train the query behavior aggressively, but keep document embeddings reusable.

\bibliography{references}

\end{document}